\documentclass[%
reprint,
nofootinbib,
 amsmath,amssymb,
 aps,
 pra,
showkeys,
]{revtex4-2}

\usepackage{mathrsfs}

\usepackage{graphicx}
\usepackage{dcolumn}
\usepackage{bm}

\usepackage{amsthm}
\usepackage[mathcal]{euscript}
\usepackage{tikz-cd}

\usepackage{url}
\usepackage{hyperref}
\usepackage{color}
\definecolor{refcolor}{RGB}{0,0,190}
\hypersetup{
    colorlinks,
    citecolor=refcolor,
    filecolor=refcolor,
    linkcolor=refcolor,
    urlcolor=refcolor
}

\usepackage{booktabs}

\usepackage{bookmark}
\bookmarksetup{
  numbered, 
  open,
}

\theoremstyle{remark}
\newtheorem{remark}{Remark}

\theoremstyle{definition}
\newtheorem{observation}{Observation}

\newtheorem{definition}{Definition}

\newtheorem{thumbrule}{Rule of Thumb}
\newtheorem{keyobservation}{Key observation}
\newtheorem{principle}{Principle}

\newtheorem{tentativePostulate}{Tentative Postulate}

\newtheorem{qmProblem}{QM Problem}
\newtheorem{mwiProblem}{MWI Problem}
\newtheorem{mstiAnswer}{MSTI Answer}

\theoremstyle{definition}
\newtheorem{defCustom}{Definition}
\renewcommand{\thedefCustom}{\arabic{definition}}
\makeatletter
\newcommand{\setdefCustomtag}[1]{
  \let\oldthedefCustom\thedefCustom
  \renewcommand{\thedefCustom}{#1}
  \g@addto@macro\enddefCustom{
    \global\let\thedefCustom\oldthedefCustom}
  }
\makeatother

\newtheorem{thesisCustom}{Thesis}
\renewcommand{\thethesisCustom}{\arabic{thesis}}
\makeatletter
\newcommand{\setthesisCustomtag}[1]{
  \let\oldthethesisCustom\thethesisCustom
  \renewcommand{\thethesisCustom}{#1}
  \g@addto@macro\endthesisCustom{
    \global\let\thethesisCustom\oldthethesisCustom}
  }
\makeatother

\newtheorem{condition}{Condition}
\renewcommand{\thecondition}{\arabic{condition}}
\makeatletter
\newcommand{\setconditiontag}[1]{
  \let\oldthecondition\thecondition
  \renewcommand{\thecondition}{#1}
  \g@addto@macro\endcondition{
    \global\let\thecondition\oldthecondition}
  }
\makeatother

\newtheorem{assumption}{Assumption}
\renewcommand{\theassumption}{\arabic{assumption}}
\makeatletter
\newcommand{\setassumptiontag}[1]{
  \let\oldtheassumption\theassumption
  \renewcommand{\theassumption}{#1}
  \g@addto@macro\endassumption{
    \global\let\theassumption\oldtheassumption}
  }
\makeatother

\newtheorem{claim}{Claim}
\renewcommand{\theclaim}{\arabic{claim}}
\makeatletter
\newcommand{\setclaimtag}[1]{
  \let\oldtheclaim\theclaim
  \renewcommand{\theclaim}{#1}
  \g@addto@macro\endclaim{
    \global\let\theclaim\oldtheclaim}
  }
\makeatother

\theoremstyle{remark}
\newtheorem{pointItem}{}
\renewcommand{\thepointItem}{\quad\arabic{pointItem}}
\makeatletter
\newcommand{\setpointItemtag}[1]{
  \let\oldthepointItem\thepointItem
  \renewcommand{\thepointItem}{#1}
  \g@addto@macro\endpointItem{
    \global\let\thepointItem\oldthepointItem}
  }
\makeatother

\begin{document}


\def\bibsection{\section*{\refname}} 
\newcommand{\pbref}[1]{\ref{#1} (\nameref*{#1})}
   
\def\({\left(}
\def\){\right)}

\newcommand{\tn}{\textnormal}
\newcommand{\ds}{\displaystyle}
\newcommand{\dsfrac}[2]{\displaystyle{\frac{#1}{#2}}}

\newcommand{\boplus}{\textstyle{\bigoplus}}
\newcommand{\botimes}{\textstyle{\bigotimes}}
\newcommand{\bcup}{\textstyle{\bigcup}}
\newcommand{\bsqcup}{\textstyle{\bigsqcup}}
\newcommand{\bcap}{\textstyle{\bigcap}}

\newcommand{\struct}{\mc{S}}
\newcommand{\kind}{\mc{K}}

\newcommand{\dddots}{\rotatebox[origin=t]{135}{$\cdots$}}

\newcommand{\statespace}{\mathcal{S}}
\newcommand{\hilbert}{\mathcal{H}}
\newcommand{\vectorspace}{\mathcal{V}}
\newcommand{\mc}[1]{\mathcal{#1}}
\newcommand{\ms}[1]{\mathscr{#1}}
\newcommand{\mf}[1]{\mathfrak{#1}}
\newcommand{\dU}{\wh{\mc{U}}}

\newcommand{\wh}[1]{\widehat{#1}}
\newcommand{\dwh}[1]{\wh{\rule{0ex}{1.3ex}\smash{\wh{\hfill{#1}\,}}}}

\newcommand{\wt}[1]{\widetilde{#1}}
\newcommand{\wht}[1]{\widehat{\widetilde{#1}}}
\newcommand{\on}[1]{\operatorname{#1}}

\newcommand{\qmU}{$\mathscr{U}$}
\newcommand{\qmR}{$\mathscr{R}$}
\newcommand{\qmUR}{$\mathscr{UR}$}
\newcommand{\qmDR}{$\mathscr{DR}$}

\newcommand{\R}{\mathbb{R}}
\newcommand{\C}{\mathbb{C}}
\newcommand{\Z}{\mathbb{Z}}
\newcommand{\K}{\mathbb{K}}
\newcommand{\N}{\mathbb{N}}
\newcommand{\Prj}{\mathcal{P}}
\newcommand{\abs}[1]{|#1|}

\newcommand{\de}{\operatorname{d}}
\newcommand{\De}{{\scriptstyle \mc{D}}}
\newcommand{\tr}{\operatorname{tr}}
\newcommand{\im}{\operatorname{Im}}

\newcommand{\ie}{\textit{i.e.}\ }
\newcommand{\vs}{\textit{vs.}\ }
\newcommand{\eg}{\textit{e.g.}\ }
\newcommand{\cf}{\textit{cf.}\ }
\newcommand{\etc}{\textit{etc}}
\newcommand{\etal}{\textit{et al.}}

\newcommand{\Span}{\tn{span}}
\newcommand{\pde}{PDE}
\newcommand{\U}{\tn{U}}
\newcommand{\SU}{\tn{SU}}
\newcommand{\GL}{\tn{GL}}
\newcommand{\su}{\tn{su}}

\newcommand{\schrod}{Schr\"odinger}
\newcommand{\vonneum}{Liouville-von Neumann}
\newcommand{\ks}{Kochen-Specker}
\newcommand{\leggarg}{Leggett-Garg}
\newcommand{\bra}[1]{\langle#1|}
\newcommand{\ket}[1]{|#1\rangle}
\newcommand{\kett}[1]{|\!\!|#1\rangle\!\!\rangle}
\newcommand{\proj}[1]{\ket{#1}\bra{#1}}
\newcommand{\braket}[2]{\langle#1|#2\rangle}
\newcommand{\ketbra}[2]{|#1\rangle\langle#2|}
\newcommand{\expectation}[1]{\langle#1\rangle}
\newcommand{\Herm}{\tn{Herm}}
\newcommand{\Sym}[1]{\tn{Sym}_{#1}}
\newcommand{\meanvalue}[2]{\langle{#1}\rangle_{#2}}
\newcommand{\Prob}{\tn{Prob}}
\newcommand{\kjj}[3]{#1\!:\!#2,#3}
\newcommand{\jk}[2]{#1,#2}
\newcommand{\JK}{\mf{j}}
\newcommand{\pobs}[1]{\mathsf{#1}}
\newcommand{\obs}[1]{\wh{\pobs{#1}}}
\newcommand{\uop}[1]{\wh{\mathbf{#1}}}

\newcommand{\weightU}[5]{\left[{#2}{}_{#3}\overset{#1}{\rightarrow}{#4}{}_{#5}\right]}
\newcommand{\weightUT}[8]{\left[{#3}{}_{#4}\overset{#1}{\rightarrow}{#5}{}_{#6}\overset{#2}{\rightarrow}{#7}{}_{#8}\right]}
\newcommand{\weight}[4]{\weightU{}{#1}{#2}{#3}{#4}}
\newcommand{\weightT}[6]{\weightUT{}{}{#1}{#2}{#3}{#4}{#5}{#6}}

\newcommand{\btimes}{\boxtimes}
\newcommand{\btimess}{{\boxtimes_s}}

\newcommand{\h}{\mathbf{(2\pi\hbar)}}
\newcommand{\x}{\boldsymbol{x}}
\newcommand{\y}{\boldsymbol{y}}
\newcommand{\z}{\mathbf{z}}
\newcommand{\q}{\mathbf{q}}
\newcommand{\p}{\mathbf{p}}
\newcommand{\0}{\mathbf{0}}
\newcommand{\annih}{\widehat{\mathbf{a}}}

\newcommand{\cs}{\mathscr{C}}
\newcommand{\ps}{\mathscr{P}}
\newcommand{\xhat}{\widehat{\x}}
\newcommand{\phat}{\widehat{\mathbf{p}}}
\newcommand{\fqproj}[1]{\Pi_{#1}}
\newcommand{\cqproj}[1]{\wh{\Pi}_{#1}}
\newcommand{\cproj}[1]{\wh{\Pi}^{\perp}_{#1}}

\newcommand{\M}{\mathbb{E}_3}
\newcommand{\D}{\mathbf{D}}
\newcommand{\dn}{\tn{d}}
\newcommand{\db}{\mathbf{d}}
\newcommand{\n}{\mathbf{n}}
\newcommand{\m}{\mathbf{m}}
\newcommand{\V}[1]{\mathbb{V}_{#1}}
\newcommand{\F}[1]{\mathcal{F}_{#1}}
\newcommand{\Fvacuumfield}{\widetilde{\mathcal{F}}^0}
\newcommand{\nD}[1]{|{#1}|}
\newcommand{\Lin}{\mathcal{L}}
\newcommand{\End}{\tn{End}}
\newcommand{\vbundle}[4]{{#1}\to {#2} \stackrel{\pi_{#3}}{\to} {#4}}
\newcommand{\vbundlex}[1]{\vbundle{V_{#1}}{E_{#1}}{#1}{M_{#1}}}
\newcommand{\rep}{\rho_{\scriptscriptstyle\btimes}}

\newcommand{\intl}[1]{\int\limits_{#1}}

\newcommand{\moyalBracket}[1]{\{\mskip-5mu\{#1\}\mskip-5mu\}}

\newcommand{\Hint}{H_{\tn{int}}}

\newcommand{\quot}[1]{``#1''}

\def\sref #1{\S\ref{#1}}

\newcommand{\dBB}{de Broglie--Bohm}
\newcommand{\dBBt}{{\dBB} theory}
\newcommand{\pwt}{pilot-wave theory}
\newcommand{\PWT}{PWT}
\newcommand{\NRQM}{{\textbf{NRQM}}}

\setlength{\belowcaptionskip}{-10pt} 
\newcommand{\image}[3]{
\begin{figure}[!ht]
\centering
\includegraphics[width=#2\textwidth]{#1}
\caption{\label{#1}#3}
\end{figure}
}

\newcommand{\TQS}{\ref{def:TQS}}
\newcommand{\HSF}{\ref{thesis:HSF}}
\newcommand{\MQS}{\ref{def:MQS}}

\title{Background freedom leads to many-worlds with local beables and probabilities}

\author{Ovidiu Cristinel Stoica}
\affiliation{
 Dept. of Theoretical Physics, NIPNE---HH, Bucharest, Romania. \\
	Email: \href{mailto:cristi.stoica@theory.nipne.ro}{cristi.stoica@theory.nipne.ro},  \href{mailto:holotronix@gmail.com}{holotronix@gmail.com}
	}%

\date{\today}

\begin{abstract}
I argue that background freedom in quantum gravity automatically leads to a dissociation of the quantum state into states having a classical space. That is, interference is not completely well-defined for states with different space geometries, even if their linear combination is.

Interference of states with different space geometry is still allowed at small scales, but precluded at macro-scales. Macrostates, including measuring devices, appear classical.

The distribution of space geometries automatically gives the Born rule.

The dissociation entails a kind of absolute decoherence, making the ad-hoc wavefunction collapse unnecessary. This naturally leads to a new version of the many-worlds interpretation, in which:

1) the classical space-states form an absolute preferred basis,

2) at any time, the resulting micro-branches are like classical worlds, with objects in space,

3) macro-branches stop interfering, even though micro-branches can interfere (as they should),

4) the space geometries converge at the Big-Bang, favoring macro-branching towards the future,

5) the wavefunctional becomes real by absorbing the phases in the global U(1) gauge,

6) ontologically, the wavefunctional consists of many gauged space-states, each of them counting as a world by having local beables (the space geometry and the classical fields),

7) the density of the classical space-states automatically obeys the Born rule.
\end{abstract}


\maketitle

\section{Introduction}
\label{s:intro}

I argue that background-free approaches to quantum gravity prevent most quantum state vectors from having physically meaningful superpositions. Interference effects require a way to relate the positions in space among different state vectors, but background freedom limits this possibility. Linear combinations exist mathematically, but interference effects are suppressed in most situations.

This leads to a new explanation of the emergence of classicality at the macro level, and to a natural derivation of the Born rule as classical probabilities of states with definite classical space.
The resulting approach to understand quantum mechanics works less naturally with the wavefunction collapse, and it implies the existence of many worlds before, but also after the wavefunction collapse. Therefore, a version of the many-worlds interpretation results, which avoids or the main problems of the standard many-worlds interpretation or provides less radical solutions.

In Sec. \sref{s:quantum-gravity} I sketch the generic features of wavefunctional formulations of background-free quantum gravity.
This leads to the notion of classical space-states, having a definite classical space (or other structure assumed to be more fundamental than the $3$d manifold).

In Sec. \sref{s:dissociation} I explain how background freedom makes the state vector dissociate into classical space-states, by limiting their ability to interfere.

In Sec. \sref{s:probabilities} I show how the distribution of space-states into which the state vector dissociates gives the Born rule. The wavefunctional contains many space-states, and the probability measure of self-location in them gives the Born rule. The wavefunctional can be made real, by absorbing the phases into the $\U(1)$ gauge of the classical fields defining the space-states. 

In Sec. \sref{s:collapse-or-many-worlds} I argue that the space-states approach works less well with the collapse postulate, but it works naturally with the many-worlds interpretation, resulting in a version of it named here the \emph{many-spacetimes interpretation} of quantum mechanics.

In Sec. \sref{s:mwi} I explain how this contributes to solving some of the main problems of the many-worlds interpretation (Fig. \ref{MSTIQM.pdf}), such as the existence of a preferred basis, the emergence of quasi-classical macro-worlds, the existence of classical-like objects in space, the time-asymmetry of the branching structure, classical probabilities, the appearance of complex numbers in quantum mechanics, and the ontology, including the local beables, which justify counting each space-state as a world.

Sec. \sref{s:discussion} concludes the article with a discussion. 

\image{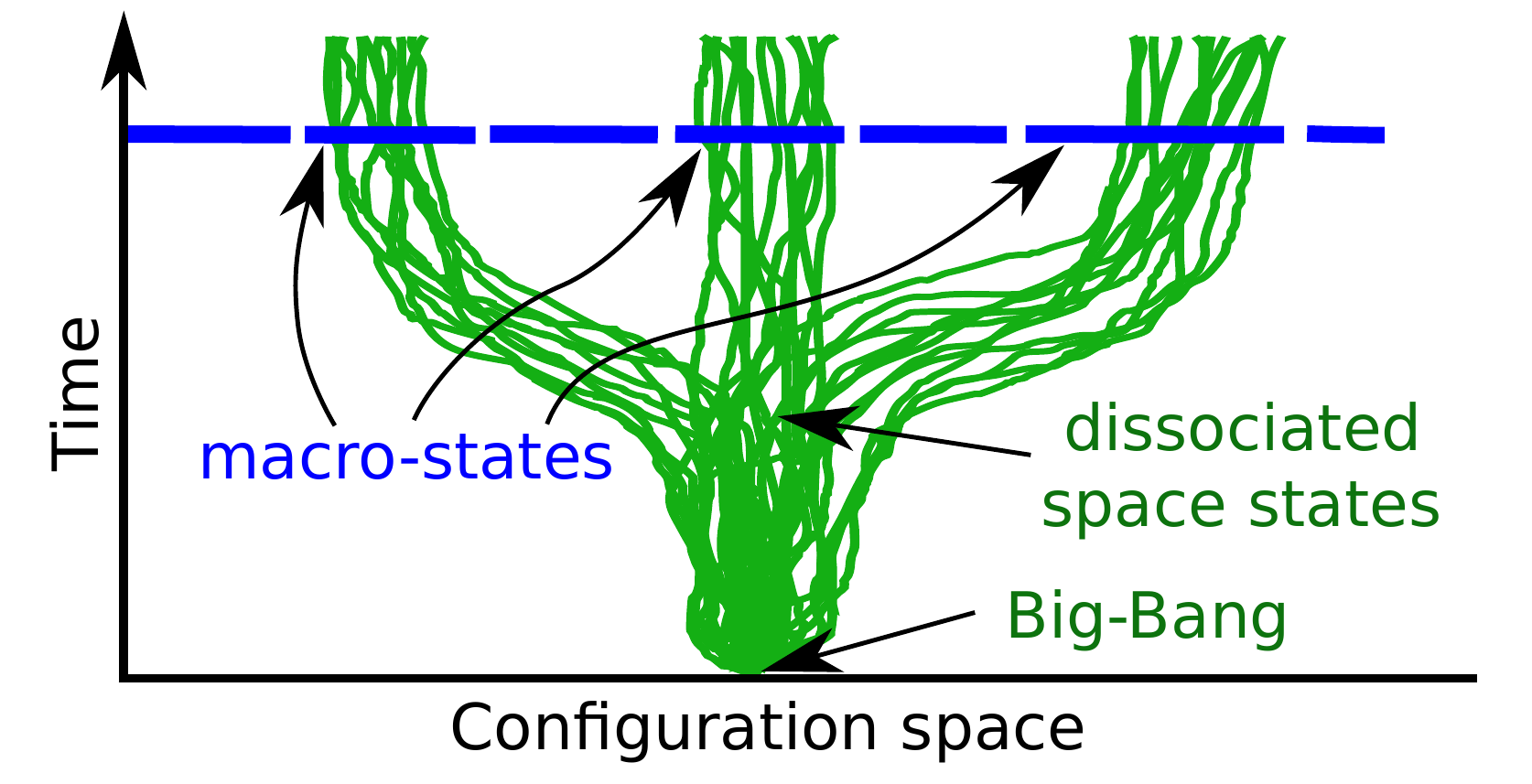}{0.45}{\textbf{Wavefunctional dissociation due to quantum-gravitational background freedom.} The microstates (in green) are space-states. They are very similar at the Big-Bang, then background freedom makes them dissociate and form a branching structure like in the many-worlds interpretation. The dissociation is reversible at micro scales, allowing interference, but it becomes irreversible when it manifests at macro scales. The branching structure (in yellow) corresponds to the macrostates (in blue). The distribution of space-states per macrostate or branch gives the Born rule.}

\section{Space-states in quantum gravity}
\label{s:quantum-gravity}

\subsection{Classical space-states}
\label{s:space}

We do not have yet a final theory of quantum gravity, and even less so one that includes the other fields.
But the universe is quantum and gravity exists, so I will assume that such a theory exists.

Many of the currently known approaches to quantum gravity admit {\schrod} wavefunctional formulations.

The Wheeler-de Witt equation
\begin{equation}
\label{eq:WdW}
\obs{H}\underline{\Psi}=0
\end{equation}
involves a wavefunctional $\underline{\Psi}=\underline{\Psi}[\gamma_{ab}]$ on the space $\tn{Riem}(\Sigma)$ of all possible Riemannian geometries $(\Sigma,\gamma_{ab})$, where $\gamma_{ab}$ is the intrinsic metric tensor on a three-dimensional manifold $\Sigma$. 
Equation \eqref{eq:WdW} was obtained \cite{Dewitt1967QuantumTheoryOfGravityI_TheCanonicalTheory} by quantizing the Hamiltonian formulation of classical general relativity by Arnowitt, Deser, and Misner (ADM) \cite{ADM1962TheDynamicsOfGeneralRelativity},
where $(\Sigma,\gamma_{ab})$ is a spacelike slice of the spacetime manifold $M=(\Sigma\times\R,g_{\mu\nu})$.

The quantization replaces the classical $3$d metric $\gamma_{ab}$ and its conjugate momentum $\pi_{\gamma}^{cd}$ with operators,
\begin{equation}
\label{eq:WdW-hp}
\begin{cases}
\wh{\gamma}_{ab}(\x)\underline{\Psi}[\gamma_{ab}]=\gamma_{ab}(\x)\underline{\Psi}[\gamma_{ab}],\\
\wh{\pi}_{\gamma}^{cd}(\x)\underline{\Psi}[\gamma_{ab}]=\dsfrac{\hbar}{i}\dsfrac{\delta\underline{\Psi}[\gamma_{ab}]}{\delta \gamma_{cd}(\x)},\\
\end{cases}
\end{equation}
subject to the canonical commutation relations
\begin{equation}
\label{eq:WdW-CCR}
\begin{cases}
\left[\wh{\gamma}_{ab}(\x),\wh{\pi}_{\gamma}^{cd}(\y)\right]=i\hbar\delta^c_{(a}\delta^d_{b)}(\x,\y),\\
\left[\wh{\gamma}_{ab}(\x),\wh{\gamma}_{cd}(\y)\right]=\left[\wh{\pi}_{\gamma}^{ab}(\x),\wh{\pi}_{\gamma}^{cd}(\y)\right]=0,\\
\end{cases}
\end{equation}
where $\x,\y\in\Sigma$ and $\delta/\delta \gamma_{cd}(\x)$ is the functional derivative.

The Wheeler-de Witt equation is a constraint equation, not an evolution equation, despite de Witt initially calling it the \emph{Einstein-{\schrod} equation}. It is complemented by three other constraint equations that factor out the space diffeomorphisms.
The wavefunctional $\underline{\Psi}$ is a timeless solution. A proposal to decode a dynamical solution, made by Page and Wootters \cite{PageWootters1983EvolutionWithoutEvolution}, consists of interpreting $\underline{\Psi}$ as a quantum system $\ket{\psi(\tau)}$ entangled with a clock $\ket{\tau}$, $\underline{\Psi}=\int_\R\ket{\tau}\ket{\psi(\tau)}d \tau$. 
This, and other proposals to recover a dynamical solution, were assessed critically in \cite{Isham1993CanonicalQuantumGravityAndTheProblemOfTime,Kuchavr2011TimeAndInterpretationsOfQuantumGravity}.
According to Page and Wootters, we can consider that the state of the universe at the time $t$ is represented by the vector $\Psi(t):=\ket{t}\ket{\psi(t)}$.

In the following we will assume the existence of a quantum theory of gravity based on time-dependent states.

Ashtekar's Hamiltonian formulation of classical general relativity \cite{Ashtekar1986NewVariablesForClassicalAndQuantumGravity} is similar to the ADM formulation, except that instead of $\gamma$ and $\pi_{\gamma}$, its variables are an $\su(2)$ connection, whose conjugate variable is a densitized frame field on $\Sigma$. 
At the classical level the ADM formalism and the Ashtekar variables are equivalent.
When quantized, the resulting operators satisfy commutation relations similar to \eqref{eq:WdW-CCR} \cite{Kiefer2012QuantumGravity}.
Its quantization was interpreted by Rovelli and Smolin in terms of \emph{loop variables} \cite{RovelliSmolin1990LoopSpaceRepresentationOfQuantumGeneralRelativity}.
We do not know if the correct formulation is one of these two or another one.

We do not even know with certainty that spacetime is continuous.
Various discrete approaches to quantum gravity are based on structures that can be represented as graphs or hypergraphs that may have attached numbers at their vertices and (hyper-)edges. For example, in the \emph{causal sets} approach \cite{Sorkin1990SpacetimeAndCausalSets}, the vertices of the graph represent \emph{events} (points from the spacetime manifold), and oriented edges join pairs of events in causal relation, in the sense that the first event is in the past lightcone of the second one.
The \emph{Regge calculus} \cite{Regge1961GeneralRelativityWithoutCoordinates} is based on triangulations of spacetime into $4$-simplices further approximated as flat. Distances are attached to the edges, and the spacetime curvature is concentrated on the $2$-faces, and it is expressed in terms of deficit angles \etc.
The \emph{causal dynamical triangulation} approach is similar, but with fixed-length edges \cite{Loll2019QuantumGravityFromCausalDynamicalTriangulationsAReview}.
\emph{Loop quantum gravity} can be formulated in terms of spin networks and spin foams. 
\emph{Spin networks} are graphs with the edges labeled by half-integer numbers corresponding to irreducible representations of the Lie algebra $\su(2)$ \cite{Penrose1971AngularMomentumAnApproachToCombinatorialSpaceTime,RovelliSmolin1995SpinNetworksAndQuantumGravity,AshtekarBianchi2021AShortReviewOfLoopQuantumGravity}. Two spin networks at different times are joined by a \emph{spin foam}, a hypergraph used in the path integral formulation of the theory.

In all these formulations, the graph or hypergraph structures are background-independent.
They can be seen as equivalence classes of (hyper)graphs embedded in the space $\Sigma$ or in the spacetime $M$, where two such embedded structures are equivalent if a diffeomorphism of the background manifold can transform one into another.
But they can also be described without being embedded in a manifold.

Many of these discrete approaches use Feynman's path integral quantization, but since the path integral associates a complex coefficient to each classical basis state, they admit a {\schrod} representation too.

Since these formulations of quantum gravity admit a wavefunctional representation, I will assume that such a formulation exists, even if we do not know yet which one is the right one.

Let $\mc{C}_S$ be the set of classical space configurations. These may be the diffeomorphism equivalence classes of Riemannian geometries $(\Sigma,\gamma)$, or more fundamental structures approximated by such geometries at low energies.
For example, if quantum gravity is one of the discrete theories whose classical configurations are labeled (hyper)graphs, these will be the elements of $\mc{C}_S$.

Most results in this article apply to both continuous and discrete spacetimes, although we will see that continuous spacetimes have some advantages.

I will assume that there is a {\schrod} formulation of quantum gravity in terms of wavefunctionals over $\mc{C}_S$, where $\mc{C}_S$ is endowed with a measure $\mu_S$.
The states of the universe are represented by unit vectors $\Psi$  in the Hilbert space $\hilbert_S$ spanned by states $\ket{\gamma}$, where $\gamma\in\mc{C}_S$ stands for $(\Sigma,\gamma_{ab})$ or the (hyper-)graph structure if the underlying quantum gravity theory is based on such a structure, with the Hermitian scalar product
\begin{equation}
\label{eq:scalar_prod_space}
\braket{\Psi}{\Psi'}:=\int_{\mc{C}_S}\Psi^\ast[\gamma]\Psi'[\gamma]\De\mu_S[\gamma].
\end{equation}

For matter quantum fields I will assume, like in the quantum field theory on the Minkowski spacetime, that there is a formulation in terms of wavefunctionals on the classical configuration space of classical fields on $\Sigma$. The classical fields include bosonic fields, which commute, and fermionic fields, which are expressed using Grassmann numbers because they anticommute at equal times, see \eg \cite{Jackiw1988AnalysisInfDimManifoldsSchrodingerRepresentationForQuantizedFields,Hatfield2018QuantumFieldTheoryOfPointParticlesAndStrings}. All other variables needed to specify how the $3$d geometries integrate into $4$d manifolds, for example the shift and lapse variables in the ADM formulation, are considered to be included as well in the configuration space of classical fields.
If space is a (hyper)graph $\gamma\in\mc{C}_S$, I assume that matter can be described, in principle, by attaching various quantities or other mathematical structures to the elements of the (hyper)graph $\gamma$.

Although ultimately the quantum field equations are those to be obeyed, each classical matter field has to be defined on an underlying classical space $\gamma$.
To express this dependence, I will denote classical matter fields by $\phi_{\gamma}$. In quantum field theory on flat space the Fock basis consists of linear combinations of classical states $\ket{\phi}$, where $\phi$ is a classical field on the space $\R^3$ \cite{Hatfield2018QuantumFieldTheoryOfPointParticlesAndStrings}.
Let us assume, for each fixed $\gamma\in\mc{C}_S$, the choice of such a basis $(\ket{\phi_{\gamma}})_{\phi_{\gamma}\in\mc{C}_{M}^{\gamma}}$, where the elements of $\mc{C}_{M}^{\gamma}$ label the basis in a way dependent on $\gamma\in\mc{C}_S$.

Let us summarize all of the above into the following
\begin{assumption}
\label{as:matter-fields}
The complete state of the universe is represented by a wavefunctional on a configuration space
\begin{equation}
\label{eq:wtC}
\mc{C}:=\bigcup_{\gamma\in\mc{C}_S}\{\gamma\}\times\mc{C}_{M}^{\gamma},
\end{equation}
endowed with the measure $\mu$ related to the measures $\mu_S$ on $\mc{C}_S$ and $\mu_M^{\gamma}$ on $\mc{C}_M^{\gamma}$ by $\De\mu[\gamma,\phi_{\gamma}]=\De\mu_S[\gamma]\De\mu_M^{\gamma}[\phi_{\gamma}]$.
Because we integrate wavefunctionals with respect to the measure $\mu$, we use the notation $\De\mu$ instead of $d\mu$.
Let the Hilbert space of such wavefunctionals be $\hilbert\cong\hilbert_S\otimes\hilbert_M$, with the Hermitian scalar product
\begin{equation}
\label{eq:scalar_prod}
\braket{\Psi}{\Psi'}:=\int_{\mc{C}}\Psi^\ast[\gamma,\phi_{\gamma}]\Psi'[\gamma,\phi_{\gamma}]\De\mu[\gamma,\phi_{\gamma}].
\end{equation}
\end{assumption}

If the manifold $\mc{C}$ would be infinite-dimensional, no Lebesgue measure $\mu$ could be defined on it. But the classical fields and metrics are constrained by the classical equations, and this severely restricts the dimension of $\mc{C}$. 
In addition, the diffeomorphism and gauge degrees of freedom are also factored out.
There are also physical reasons to assume that $\dim\mc{C}<\infty$, due to the \emph{entropy bound} \cite{Bekenstein1981UniversalUpperBound,Bekenstein2005HowDoesEntropyInformationBoundWork}, and also severe constraints related to the arrow of time \cite{Stoica2022DoesQuantumMechanicsRequireConspiracy}. So I will assume that $\mc{C}$ admits a measure $\mu$.

\subsection{Macrostates and ontic microstates}
\label{s:ontic-micro-macrostates}

Macrostates represent the states observable at the macroscopic level. Because by knowing the macrostate we can only know with limited certainty the true state (named in this context ``microstate''), macrostates correspond to equivalence classes of microstates that cannot be distinguished macroscopically, by direct observation.
Since different macrostates are distinguishable, there is a set of commuting projectors $(\obs{P}_{\alpha})_{\alpha\in\mc{A}}$ on $\hilbert$, so that $[\obs{P}_{\alpha},\obs{P}_{\beta}]=0$ for any $\alpha\neq\beta\in\mc{A}$, and $\boplus_{\alpha\in\mc{A}}\obs{P}_{\alpha}\hilbert=\hilbert$.
Any \emph{macrostate} is represented by a subspace of the form $\obs{P}_{\alpha}\hilbert$. We call \emph{quasi-classical} the states belonging to macrostates $\obs{P}_{\alpha}\hilbert$.

Since all projectors $\obs{P}_{\alpha}$ commute, there are bases of $\hilbert$ consisting of common eigenvectors of $(\obs{P}_{\alpha})_{\alpha\in\mc{A}}$. 
Since both $\gamma$ and $\phi_{\gamma}$ are classical, it makes sense to assume that the states $\ket{\gamma,\phi_{\gamma}}$ are quasi-classical, \ie $\ket{\gamma,\phi_{\gamma}}\in\obs{P}_{\alpha}\hilbert$ for some $\alpha$.
This expresses the following idea:
\begin{observation}[Macro classicality]
\label{obs:macro-classicality}
At the macro level, the world looks like the classical world.
\end{observation}

This justifies the following

\begin{assumption}
\label{as:quasi-classical}
The basis $(\ket{\gamma,\phi_{\gamma}})_{(\gamma,\phi_{\gamma})\in\mc{C}}$ consists of quasi-classical states, \ie for any $\ket{\gamma,\phi_{\gamma}}$ there is an $\alpha\in\mc{A}$ so that $\ket{\gamma,\phi_{\gamma}}\in\obs{P}_{\alpha}\hilbert$.
\end{assumption}

The existence of a basis of quasi-classical states is guaranteed by the existence of the macro-projectors. 
Assumption \ref{as:quasi-classical} specifies that such a basis can consist of states with classical geometry and classical matter fields.

\begin{definition}
\label{def:space-state}
States of the form $\ket{\gamma,\phi_{\gamma}}$ will be called \emph{space-states}, and also \emph{ontic states} for reasons that will be explained later.
\end{definition}

We expect that a space-state immediately evolves into a linear combination of space-states.
Dissociation and reassociation happen continuously.
However, at the macro level, the state may remain quasi-classical under unitary evolution for longer time intervals. This accounts for the fact that macroscopic systems do not evolve all the time into linear combinations of macrostates like the {\schrod} cat, although unitary evolution may lead to such linear combinations during quantum measurements.

\subsection{Space-states are fundamental}
\label{s:fundamental-space-states}

Just because physicists first discovered classical physics, and later quantum theory, and formulated the latter by quantizing the former, it does not mean that quantum theory requires classical physics to exist. The universe is what it is, and it is fundamentally quantum.

However, the Hilbert space is too symmetric as it is, and without the existence of preferred structures that break its symmetry, there would be no relation between Hilbert space vectors and physical reality, nor between Hermitian operators and physical observables. Physical properties cannot simply emerge from the abstract state vector, even if the Hamiltonian is known, because if they would, infinitely many entities with the very same properties, but able to represent completely different physical realities, would emerge as well \cite{Stoica2021SpaceThePreferredBasisCannotUniquelyEmergeFromTheQuantumStructure,Stoica2022VersatilityOfTranslations}.
Therefore, the basis $(\ket{\gamma,\phi_{\gamma}})_{(\gamma,\phi_{\gamma})\in\mc{C}}$ is special among the others, because of its physical meaning.
This justifies

\begin{assumption}
\label{as:fundamental-space-states}
The space-states are fundamental, in the sense that, by their physical meaning, they are special among the other states represented by $\hilbert$.
\end{assumption}

As explained earlier, the states $\gamma$ are not necessarily Riemannian geometries, they can be other structures approximated at low energies by such geometries.
What is important is that they have a special physical meaning, in the same sense in which, in nonrelativistic quantum mechanics, the position operators and their eigenvectors have a special physical meaning compared to other operators or vectors in the Hilbert space.
Similar considerations hold for the matter fields.

\section{Dissociation into space-states}
\label{s:dissociation}

\subsection{Background freedom}
\label{s:background-freedom}

To construct the configuration space $\mc{C}$, we eliminated the unphysical degrees of freedom due to diffeomorphisms and global gauge transformations.
For example, two metric tensor fields on $\Sigma$ may look different, but a coordinate transformation, which corresponds to a diffeomorphism of $\Sigma$, may be able to map them into one another, in which case they are isometric.
For this reason, we took as classical states the equivalence classes of metrics on $\Sigma$ under diffeomorphisms.
Matter fields also have ``internal background freedom'' (gauge freedom).

Similarly if space is a discrete structure like the ones that can be represented by graphs or hypergraphs from \sref{s:space}, the configuration space consists of such structures based on their internal relations, not as particular embeddings in a $3$d manifold.
But let us state this explicitly, since it will be central in the article:
\begin{assumption}
\label{as:background-freedom}
Our theory is background-free.
\end{assumption}

All approaches to quantum gravity mentioned in Sec. \sref{s:quantum-gravity} are background-free.
The case for background freedom was made for example by Smolin \cite{Smolin2006TheCaseForBackgroundIndependence}.
General relativity already shows that the structures have to be relational: we use coordinates, but they are not absolute, they are just ways to assign numbers to points in space or spacetime. The \emph{hole argument} \cite{Stachel2014TheHoleArgumentAndSomePhysicalAndPhilosophicalImplications,Norton1988TheHoleArgument} shows that we should not assume that the points of the underlying manifold have an independent reality from the intrinsic relations introduced by the metric tensor, because this leads to indeterminacy in a supposedly deterministic theory.

This is why many of the approaches to quantum gravity seem to require background freedom, or even have it built-in. This is true for the formulation based on the Wheeler-de Witt equation \eqref{eq:WdW}, the discrete approaches based on (hyper)graphs discussed earlier, like causal sets, Regge calculus, causal dynamical triangulations, loop quantum gravity \etc.
For a discussion of background independence in \emph{string theory} see Witten \cite{Witten1993QuantumBackgroundIndependenceInStringTheory}.

\subsection{Background freedom and dissociation}
\label{s:background-dissociation}

In general, we make no distinction between the concepts of linear combination and superposition, except maybe that a linear combination is understood as the mathematical expression of a superposition, which is a physical concept related to the position in space and phenomena like interference.
And these two terms usually coincide.

In nonrelativistic quantum mechanics, any two wavefunctions can be superposed in space, because the underlying geometry is the same, and the reference frames are the same.
In the wavefunctional formulation of quantum field theory, the local information about the wavefunctional of a scalar field is obtained by using local operators at $\x\in\Sigma=\R^3$, definable in function of the operators $\wh{\varphi}(\x)$ and $\wh{\pi}_{\varphi}(\x)$ (to be rigorous, one uses operator-valued distributions, applied to a sequence of test functions that converge uniformly to the Dirac distribution $\delta_{\x}$).

In background-dependent theories of quantum gravity we can define local operators in a similar way, in function of the operators $\wh{\gamma}(\x)$ and $\wh{\pi}_{\gamma}(\x)$, and $\wh{\varphi}_{\alpha}(\x)$ and $\wh{\pi}_{\varphi_{\alpha}}(\x)$ for each component $\varphi_{\alpha}$ of the matter field  $\varphi$, where $\alpha$ stands for the spin and the internal degrees of freedom of the field.

But in background-free quantum gravity, local operations on space, and therefore superpositions, do not always make sense, even if the linear combinations are always defined.
If the theory is background-free, a difference appears when we apply local operators to linear combinations. Any local operator $\obs{A}(\x)$ depends on $\x$, but background freedom prevents uniquely matching points from $\Sigma$ for $\ket{\gamma,\phi_{\gamma}}$ to points from $\Sigma$ for $\ket{\gamma',\phi_{\gamma'}'}$, because in general $\gamma\neq\gamma'$.
The situation is even more visible in background-free theories where $(\Sigma,\gamma)$ is replaced by a labeled (hyper)graph, because in this case correspondences between the vertices of different (hyper)graphs are not even possible in general.

A correspondence between the points of $\Sigma$ for $\ket{\gamma,\phi_{\gamma}}$ and those of $\Sigma$ for $\ket{\gamma',\phi_{\gamma'}'}$ requires $(\Sigma,\gamma)$ and $(\Sigma,\gamma')$ to be isometric. Sometimes such a correspondence exists only between some open regions of $\Sigma$. So the dissociation is not always ensured, and we will see that this is essential.
Because local operators and superpositions are not completely well-defined in the absence of a common background, we arrived at the following:
\begin{keyobservation}
\label{obs:key:dissociation}
Background freedom implies the dissociation of the wavefunctional into space-states.
\end{keyobservation}

The dissociation is not necessarily complete, and various cases are captured in the following definition.

\begin{definition}
\label{def:dissociation}
Two state vectors $\ket{\gamma}\ket{\Psi_M}$ and $\ket{\gamma'}\ket{\Psi_M'}$ are \emph{locally associable} if there exist two non-empty open subsets $U,U'\subseteq\Sigma$ and an isometry between $(U,\gamma)$ and $(U',\gamma')$. If $U=U'=\Sigma$, they are \emph{globally associable}.

Two state vectors are \emph{dissociated} if they are not globally associable.
They are \emph{partially dissociated} if they are locally but not globally associable.
They are \emph{completely dissociated} if they are not associable (globally or locally).
\end{definition}

In the discrete case, in Definition \ref{def:dissociation}, (local) isometries are replaced by (local) isomorphisms between the labeled (hyper)graphs $\gamma$ and $\gamma'$.

It is possible for space-states to reassociate, at least partially. This allows quantum interference to exist at micro scales, and it is the key to understanding why our quantum world looks quantum at small scales, and classical at macro scales.

When the system evolves into associable states, the association is allowed only if it is consistent with the dynamics, which is given by a local Hamiltonian \cite{Hatfield2018QuantumFieldTheoryOfPointParticlesAndStrings}.

It is interesting to compare dissociation induced by background freedom with Penrose's \emph{gravitational decoherence}. According to Penrose \cite{Penrose1986GravityAndStateVectorReduction}, 
``I envisage that nature feels uncomfortable about `linearly superposing' space-time geometries which differ significantly from one another and, instead, prefers to settle for (i.e. to `reduce' to) essentially just one of the geometries involved.'' Then, he proposes that this requires a nonlinear collapse due to the gravitational entropy. Later he justifies the collapse by invoking the existence of different time parameters for the superposed spacetimes \cite{Penrose1996OnGravitysRoleInQuantumStateReduction}.

By contrast, the dissociation induced by background freedom proposed in this article is simply due to the absence of a correspondence between the points in different spaces in the absence of common background. The dissociation is accepted as it is, and no change of the {\schrod} wavefunctional equation is postulated to get rid of it.

For $4$d-spacetimes, background freedom may lead to an even stronger dissociation than for spaces. However, because we start with Hamiltonian approaches to quantum gravity, and it is expected that space-states evolve into linear combinations of space-states, space-state dissociation is already significant enough.

\section{Probabilities from counting space-states}
\label{s:probabilities}

\subsection{Taking dissociation seriously}
\label{s:dissociation-seriously}

Every vector $\ket{\Psi}$ from $\hilbert$ has the form
\begin{equation}
\label{eq:state-in-basis}
\ket{\Psi} = \int_{\mc{C}}c[\gamma,\phi_{\gamma}]\ket{\gamma,\phi_{\gamma}}\De\mu[\gamma,\phi_{\gamma}],
\end{equation}
where $c[\gamma,\phi_{\gamma}]=\Psi[\gamma,\phi_{\gamma}]\in\C$, $\gamma\in\mc{C}_S$, and $\ket{\phi_{\gamma}}$ belongs to the basis $(\ket{\phi_{\gamma}})_{\phi_{\gamma}\in\mc{C}_{M}^{\gamma}}$ which may depend on the geometry $\gamma$.

We may be tempted to simply proclaim the Born rule, asserting that the probability density is
\begin{equation}
\label{eq:born_rule}
P[\gamma,\phi_{\gamma}]=\abs{c[\gamma,\phi_{\gamma}]}^2.
\end{equation}

But let us resist this for a while, and explore the consequences of the dissociation. 
If we explore the consequences of a physical principle, we should do it in its own terms, and if the result contradicts the observations, we should drop the starting principle.

The dissociation into space-states suggests:
\begin{principle}
\label{pp:dissociation}
Each space-state is either not present in $\ket{\Psi(t)}$, or it is present once (although the present space-states can be distributed with a nonuniform density).
\end{principle}

This ``either-or'' is vague at this point, but it will be clarified later in this Section. At any rate, it may seem to contradict everything we know.
However, we will get quantum theory back, with the familiar complex numbers, which will receive a geometric meaning in terms of a global gauge, and, if $\mc{C}$ is continuous, with the Born rule as we know it, but resulting from ``counting'' space-states.

\subsection{Making the wavefunctional real}
\label{s:real-wavefunctional}

Background freedom implies that the quantum state dissociates automatically into space-states, but since the coefficients $c[\gamma,\phi_{\gamma}]$ from eq. \eqref{eq:state-in-basis} are complex numbers, we need to understand their meaning.

Let us express the complex coefficients $c[\gamma,\phi_{\gamma}]$ from eq. \eqref{eq:state-in-basis} in the polar form
\begin{equation}
\label{eq:polar-form}
c[\gamma,\phi_{\gamma}]=r[\gamma,\phi_{\gamma}]e^{i \theta[\gamma,\phi_{\gamma}]},
\end{equation}
with $r[\gamma,\phi_{\gamma}]\geq 0$.

If $\phi_{\gamma}$ transforms nontrivially under a global $\U(1)$ gauge transformation, let $e^{i \theta}\phi_{\gamma}$ represent a global gauge transformation of $\phi_{\gamma}$. 
The particular form of the action of $e^{i \theta}$ on $\phi_{\gamma}$ depends on the geometric meaning of the field, but I will use a uniform notation for this action.

The classical fields $\phi_{\gamma}$ and $e^{i \theta}\phi_{\gamma}$ are physically equivalent, and we denote this by $\phi_{\gamma}\approx e^{i \theta}\phi_{\gamma}$. The state vectors $\ket{\phi_{\gamma}}$ and $e^{i \theta}\ket{\phi_{\gamma}}$ are distinct vectors, but they represent the same physical state, so we write $\ket{\phi_{\gamma}}\sim e^{i \theta}\ket{\phi_{\gamma}}$.

This suggests the following interpretation:
\begin{keyobservation}
\label{obs:key:gauge}
If the matter fields admit a nontrivial global $\U(1)$ gauge symmetry, we can make for any $\theta\in\R$ the identification
\begin{equation}
\label{eq:gauge}
e^{i \theta}\ket{\gamma,\phi_{\gamma}}\equiv\ket{\gamma,e^{i \theta}\phi_{\gamma}}.
\end{equation}
\end{keyobservation}

The identification \eqref{eq:gauge} is consistent with the fact that $e^{i \theta}\ket{\gamma,\phi_{\gamma}}$ and $\ket{\gamma,e^{i \theta}\phi_{\gamma}}$ represent the same physical state, accounting for the fact that the physical equivalence of the classical fields $\phi_{\gamma}$ and $e^{i \theta}\phi_{\gamma}$ corresponds to the physical equivalence of the state vectors $\ket{\phi_{\gamma}}$ and $e^{i \theta}\ket{\phi_{\gamma}}$.
We summarize this in the commutative diagram \eqref{eq:gauge-phase-equiv}.
\begin{equation}
\label{eq:gauge-phase-equiv}
\begin{tikzcd}[column sep=9 em,row sep=5.5 em,
 /tikz/column 2/.append style={anchor=base east}]
\phi_{\gamma} \arrow[d,"\textnormal{quantization}" {anchor=south, rotate=-90}]
\arrow[d,"\textnormal{quantization}" {anchor=south, rotate=-90},shift left=16.5em]
 \arrow[r,mapsto,"\textnormal{gauge transformation}"]
 & e^{i \theta}\phi_{\gamma}\ \, \\
\left|\phi_{\gamma}\right\rangle \arrow[r,mapsto,"\textnormal{phase transformation}"] & e^{i \theta}\left|\phi_{\gamma}\right\rangle\equiv\left|e^{i \theta}\phi_{\gamma}\right\rangle
\end{tikzcd}
\end{equation}

This works for fields that admit an $\U(1)$ symmetry, like charged fields, gauge potentials, and spinor fields, but it is sufficient that $\phi_{\gamma}$ includes one such field.

Then, eq. \eqref{eq:state-in-basis} becomes
\begin{equation}
\label{eq:state-in-basis-polar}
\ket{\Psi} = \int_{\mc{C}}r[\gamma,\phi_{\gamma}]\ket{\gamma,e^{i \theta[\gamma,\phi_{\gamma}]}\phi_{\gamma}}\De\mu[\gamma,\phi_{\gamma}].
\end{equation}

We see that each physical classical field contributes to $\ket{\Psi}$ with a uniquely determined gauge $e^{i \theta[\gamma,\phi_{\gamma}]}$ and real coefficient $r[\gamma,\phi_{\gamma}]$. As $\ket{\Psi}$ evolves in time, the gauge and $r[\gamma,\phi_{\gamma}]$ can change.

It remains to explain the relation between $r[\gamma,\phi_{\gamma}]$ and the probability density of ontic states, in agreement with Principle \ref{pp:dissociation}.

\subsection{Emergence of the Born rule}
\label{s:Born}

Now that we have seen that gauge freedom allows the coefficients in the linear combination \eqref{eq:state-in-basis} to be real numbers, let us see what their meaning is and how it relates to probabilities.

I will assume that the configuration space $\mc{C}$ is continuous.
This likely requires that $\Sigma$ is a $3$d manifold.
I show that, under this assumption, the Born rule emerges as the continuous limit of ``counting'' space-states.
In fact, since the basis is continuous, it is uncountable, so we will obtain measures, as in classical probabilities.
A more detailed discussion can be found in \cite{Stoica2022BornRuleQuantumProbabilityAsClassicalProbability}.

Let us apply the necessary gauge transformations to the fields $\phi_{\gamma}$ so that $\theta[\gamma,\phi_{\gamma}]=0$ in eq. \eqref{eq:polar-form}.

We denote for simplicity $\xi:=(\gamma,\phi_{\gamma})$.

First, we notice that a state vector of the form $
\ket{\Psi}=\frac{1}{\sqrt{n}}\sum_{k=1}^n\ket{\xi_k}$, where $(\ket{\xi_k})_{k\in\{1,\ldots,n\}}$ are distinct basis vectors, leads to the Born rule.
If $\obs{P}_{\alpha}$ is a macro-projector and $n_{\alpha}$ basis vectors composing $\ket{\Psi}$ belong to $\obs{P}_{\alpha}\hilbert$, then $\bra{\Psi}\obs{P}_{\alpha}\ket{\Psi}=n_{\alpha}/n$.
Therefore, the Born rule simply coincides with the usual counting rule ``probability is the ratio of the number of favorable outcomes to the total number of possible outcomes''.
But only a small subset of the possible state vectors have this form, so this idea fails if the basis is finite or discrete.

However, this idea works in the continuous case, since the basis vectors can be distributed with nonuniform density.
Suppose that the macro-projectors are compatible with the basis $(\ket{\xi})_{\xi\in\mc{C}}$, in the sense that they have the form
$\obs{P}_{\alpha}=\int_{\mc{C}_{\alpha}}\ket{\xi}\bra{\xi}\De\mu[\xi]$, where the set $\mc{C}_{\alpha}\subset\mc{C}$ is $\mu$-measurable.
In \cite{Stoica2022BornRuleQuantumProbabilityAsClassicalProbability} it was shown how we can construct, for any unit vector $\ket{\Psi}\in\mc{C}$, an infinite sequence $(\mc{P}_n)_{n\in\N}$ of sets of projectors so that (i) the projectors from each set $\mc{P}_n$ are compatible with the basis $(\ket{\xi})_{\xi\in\mc{C}}$, (ii) each set of projectors $\mc{P}_n$ is determined by a partition of $\mc{C}$ into measurable subsets, (iii) each set of projectors $\mc{P}_{n+1}$ refines the previous set $\mc{P}_n$, (iv) the sequence $(\mc{P}_n)_{n\in\N}$ converges to a refinement of the set of macro-projectors $(\obs{P}_{\alpha})_{\alpha\in\mc{A}}$, and (v) at each step $n$, the projectors from $\mc{P}_n$ determine a decomposition of $\ket{\Psi}$ of the form $\ket{\Psi}=\frac{1}{\sqrt{2^n}}\sum_{k=1}^{2^n}\ket{\xi_{\alpha,k}}$, where each $\ket{\xi_{\alpha,k}}$ is a unit vector.
This decomposition converges of course to $\ket{\Psi}$. Since at each step the decomposition gives the Born rule by state counting, the limits gives the Born rule.
Since the projectors in each partition $\mc{P}_n$ are compatible with the basis $(\ket{\xi})_{\xi\in\mc{C}}$, the Born rule can be understood as a probability measure over the ontic states.

But, even if the measure satisfies the Born rule, we have to interpret it probabilistically. This is achieved by the following Principle:

\begin{principle}
\label{pp::worlds}
The possible worlds correspond to space-states.
An agent being part of a world, probabilities represent the agent's ignorance of the state of the world.
\end{principle}

If we apply these results to the many-worlds interpretation, multiple space-states coexist, so the ignorance of the agent becomes ignorance about the agent's self-location in the space-states composing the state vector $\ket{\Psi}$.
If we apply them to standard quantum mechanics, one may think that probabilities represent the ignorance about the space-state, supposed to be unique, but in Sec. \sref{s:collapse-or-many-worlds} we will see that more space-states have to coexist in $\ket{\Psi}$, even if it collapses.

There is a way to interpret $\ket{\Psi}$ as a densitized set of ontic states from $\mc{C}$.
If $r[\xi]:=r[\gamma,\phi_{\gamma}]$ from eq. \eqref{eq:state-in-basis-polar} is $\mu$-measurable, we can define a new measure
\begin{equation}
\label{eq:density_measure}
\De\wt{\mu}[\xi]:=r[\xi]\De\mu[\xi],
\end{equation}
and obtain
\begin{equation}
\label{eq:psi_real_uniformized}
\ket{\Psi}=\int_{\mc{C}}\ket{\xi} \De\wt{\mu}[\xi].
\end{equation}

Since $r[\xi]$ is $\mu$-measurable, the measure $\wt{\mu}$ is absolutely continuous with respect to $\mu$.

Using the measure $\wt{\mu}$ in the uniform form of $\ket{\Psi}$ from \eqref{eq:psi_real_uniformized} gives, of course, the same norm as using the measure $\mu$ in the usual form,
\begin{equation}
\label{eq:check-normalized}
\left|\int_{\mc{C}_{\alpha}}\ket{\xi} \De\wt{\mu}[\xi]\right|^2=\bra{\Psi}\obs{P}_{\alpha}\ket{\Psi}.
\end{equation}

But, in addition, it allows the interpretation of $\ket{\Psi}$ as a densitized set of classical states. Since an agent or observer supervenes on a classical state, it makes sense to interpret the agent's self-location ignorance of the microstate on which it supervenes as a classical probability \cite{Stoica2022BornRuleQuantumProbabilityAsClassicalProbability}.
This gives the Born rule in agreement with Principle \ref{pp:dissociation} (Fig. \ref{3dspace-counting.pdf}).
The role of the new measure $\wt{\mu}$ is just to show that the states satisfy Principle \ref{pp:dissociation}, it does not change the original measure, it only expresses it differently, so that the presence of the ontic states become apparent.
A more detailed explanation can be found in \cite{Stoica2022BornRuleQuantumProbabilityAsClassicalProbability}.

Note that it is not necessary to prove the Born rule for individual particles or subsystems. Since any quantum measurement ultimately becomes an observation of the macrostate, it is sufficient to prove the Born rule for macro-branches or macro-projectors. And we have seen that the measure of micro-branches per macro-branch gives the Born rule.

\image{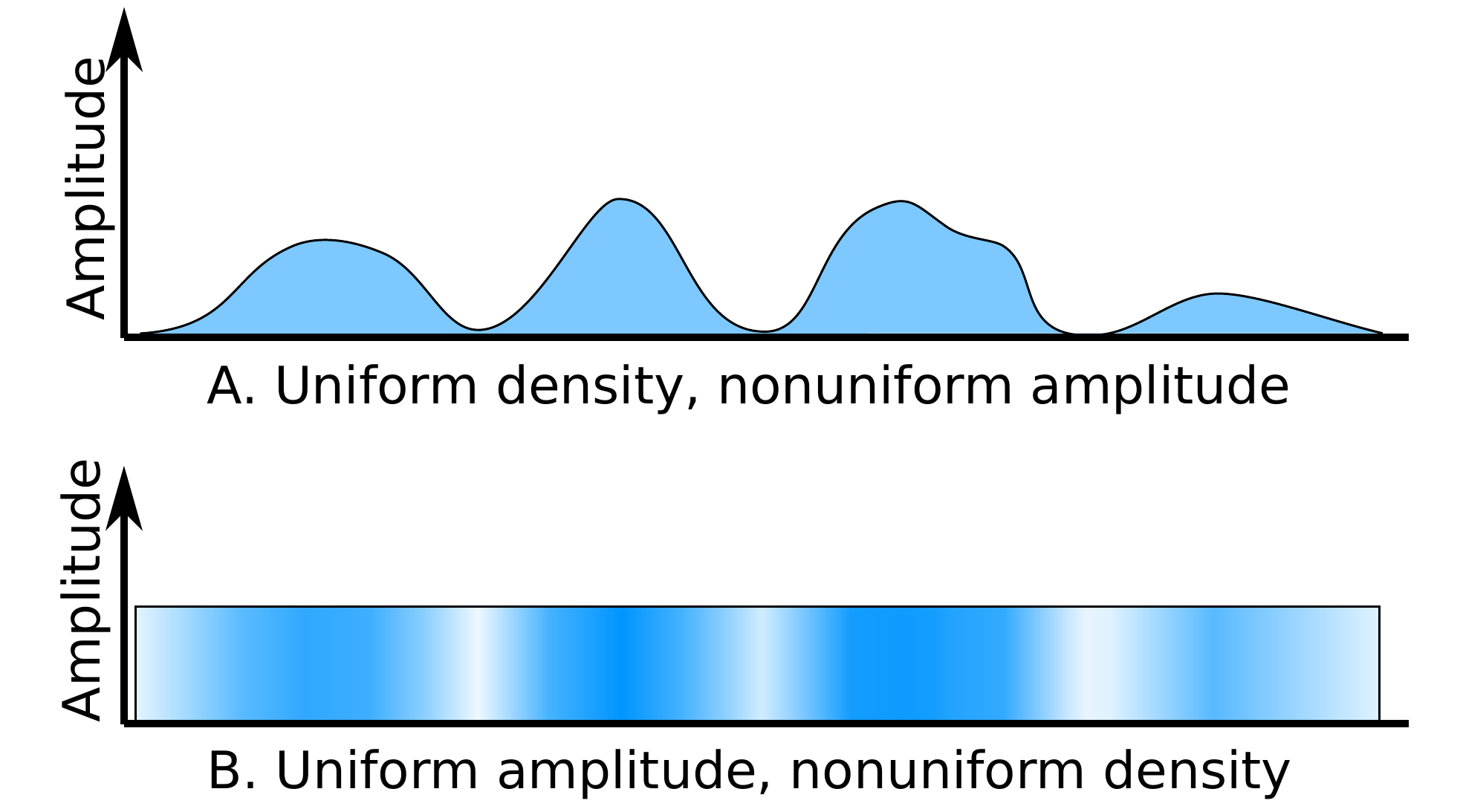}{0.45}{\textbf{The Born rule from counting space-states.}
\\\textbf{A.} The usual interpretation of a wavefunction as a linear combination of basis state vectors of different norms.
\\\textbf{B.} The interpretation of the wavefunction in terms of constant norm basis state vectors, but with inhomogeneous density.}

Therefore, the numbers from eq. \eqref{eq:state-in-basis-polar} have a direct meaning: Principle \ref{pp:dissociation} combined with the gauge freedom allows the interpretation of the states $\ket{\Psi}$ as consisting of space-states that are either present or not, and the present ones are distributed according to a density.
We obtained the Born rule from ``counting'' space-states.

\begin{keyobservation}
\label{obs:key:distribution}
If $\mc{C}$ is continuous, any state vector $\ket{\Psi}\in\hilbert$ consists of mutually orthogonal space-states distributed according to a density.
Then, for the macro-projectors $(\obs{P}_{\alpha})_{\alpha\in\mc{A}}$, the Born rule results as the probability density of space-states in the classical sense.
\end{keyobservation}

\begin{remark}
\label{rem:born_rule_more_general}
This derivation of the Born rule is not limited to the case when the basis states are space-states \cite{Stoica2022BornRuleQuantumProbabilityAsClassicalProbability}. 
What is important is that the basis is continuous, that the basis vectors belong to macrostates, and that they can support agents or observers. Only then it makes sense to talk about self-location probability.
In quantum field theory in the {\schrod} wavefunctional representation, one can use the classical field configurations to obtain the basis.
But space-states have the advantage of dissociating in a natural way, and of including gravity.
Moreover, the space-states consist of local beables, which are $\gamma$ and $\phi_{\gamma}$ (see Sec. \sref{s:ontology}).
This justifies considering these states as worlds supporting agents that can reason about the statistics of the microstates and self-location uncertainty.
\qed
\end{remark}

\section{Collapse postulate or many-worlds?}
\label{s:collapse-or-many-worlds}

Let us see how dissociation into space-states works with quantum measurements, and whether it works better by assuming the collapse postulate or with the many-worlds interpretation.

A measuring device is a quantum system in a quasi-classical state.
When interacting with the observed system, assumed to be microscopic in the sense that it is not directly observable, the combined system evolves into a linear combination of macroscopically distinct states.
Each of these states contains the observed system in a different state, and the pointer of the measuring device indicating that state.
So the {\schrod} equation predicts that two or more stories describing the measurement are simultaneously true.
But we never observe such linear combinations: after the measurement, the pointer state is always in a definite macrostate.

\begin{qmProblem}
\label{qmProblem:collapse}
Why do all linear combinations appear to be possible at micro-scales, but not at macro-scale?
\end{qmProblem}

To resolve this problem, in standard quantum mechanics one invokes the \emph{collapse postulate} \cite{vonNeumann1955MathFoundationsQM}, which simply states that quantum measurements suspend the {\schrod} evolution, so that from the linear combination we keep only the term that corresponds to one of the possible pointer states, removing the others.

In doing this, standard quantum mechanics assumes, without explaining it, the pre-existence of measuring devices in quasi-classical states. But, since most quantum states are linear combinations of quasi-classical states, we have the following problem:
\begin{qmProblem}
\label{qmProblem:measuring-device}
Why is the measuring device already in a quasi-classical state?
\end{qmProblem}

The collapse postulate purports to solve QM Problem \ref{qmProblem:collapse} by assuming implicitly that QM Problem \ref{qmProblem:measuring-device} is already solved.
However, both problems can be solved simultaneously, by extending the collapse postulate to apply not only to measurements, but whenever a quantum system becomes a linear combination of states that belong to distinct macrostates. Then, when any quantum system is no longer quasi-classical, the collapse postulate is triggered and only one of the macroscopic possibilities remains \cite{Stoica2020StandardQuantumMechanicsWithoutObservers}.

A problem with the collapse postulate is that the {\schrod} equation is considered valid in some situations, and suspended in other situations.

\emph{There seems to be a double standard here}.
On one hand, linear combinations and entangled states appear and evolve in parallel as long as no observation is made or at least the linear combinations remain in the same macrostate. The experiments are consistent with this. On the other hand, if the linear combinations extend across multiple macrostates, we allow only one of the histories, and censor all the other ones, by appealing to the collapse postulate.

We can try to use the space-states approach to solve QM Problems \ref{qmProblem:collapse} and \ref{qmProblem:measuring-device} at once, by reformulating the collapse postulate in the following way:

\begin{tentativePostulate}[Space-states Collapse Postulate]
\label{altPost:collapse}
During the evolution of the system, the space-states may become irreversibly dissociated into two or more sets of space-states, corresponding to multiple macrostates. When this happens, only one of the macrostates remains, and the others disappear.
The probability is given by the ratio between the measure of the space-states in the remaining macrostate to the total measure of space-states before the collapse.
\end{tentativePostulate}

This Tentative Postulate seems to provide a basis to explain macro systems, including measuring devices. If so, it can solve both QM Problems \ref{qmProblem:collapse} and \ref{qmProblem:measuring-device} at once.

But dissociation and reassociation happen all the time.
Reassociation allows interference effects, but when dissociation is irreversible, these effects are suppressed automatically. 
Therefore, the collapse is still arbitrary, there is still no clear rule when it should be invoked. 
When no measurement is made, multiple space-states are allowed to coexist, dissociate and associate in interference patterns in the wavefunctional. But when a measurement is made, only a subset of the space-states seem to remain, based on macroscopic criteria. Some linear combinations of space-states seem to be ``more equal'' than others.
Dissociation makes this arbitrariness more evident, because the micro-branching is well-defined by the geometry compared with the usual superpositions, yet the space-states are kept or removed as it is more convenient.

\begin{remark}
\label{rem:born_rule_collapse}
If we assume collapse and try to explain the Born rule by counting space-states as in Sec. \sref{s:probabilities}, we will have to accept that the wavefunction consists of many microstates that exist simultaneously, and part of them are eliminated by every collapse. But this would make quantum mechanics with the collapse postulate a strange version of the many-worlds interpretation, in which some of the micro-branches are removed with every collapse, and others are kept, based on purely macroscopic criteria.
Those space-states that are kept have to belong to the same macrostate, one cannot just keep space-states from different macrostates.
\qed
\end{remark}

These remarks immediately prompt the following:
\begin{keyobservation}
\label{obs:key:mwi}
Tentative Postulate \ref{altPost:collapse} is unnecessary, because once the dissociation becomes irreversible, the macro-branches evolve independently and no longer interfere, and no space-states from different macro-branches associate again.
This avoids the necessity to remove space-states based on macroscopic criteria.
\end{keyobservation}

Therefore, since once the dissociation becomes irreversible at macro scales the macro-branches no longer reassociate anyway, the space-states approach works more naturally with  the many-worlds interpretation (MWI) rather than with the wavefunction collapse.

The key idea of the MWI is to take the {\schrod} equation seriously, without introducing any ad-hoc rule that applies only to macro scales.
This implies that all possible components of the total wavefunction continue to exist after the measurement, but thanks to decoherence, they no longer ``see'' each other.
The linearity of the {\schrod} equation allows the macroscopically distinct states that result from a quantum measurement by unitary evolution to be independent, but in addition, they no longer interfere. The wavefunction is branching so that the different branches occupy different regions in the configuration space.
Interference is suppressed because the copy of any agent or measuring device in one branch is unable to detect anything from another branch, so the branches no longer ``know'' about one another.
And the branches become macroscopically distinct, in the sense that they correspond to projections of the state vector on different macrostates $\obs{P}_{\alpha_1}\hilbert,\ldots,\obs{P}_{\alpha_n}\hilbert$.

Decoherence into macro-branches seems to explain the existence of measuring devices and to solve the measurement problem without violating the {\schrod} equation by invoking an ad-hoc wavefunction collapse.

There are several problems that are not solved, at least not in a generally accepted way or so that it does not require a complete reinterpretation of well-established concepts like probabilities. They will be discussed in Sec. \sref{s:mwi}, where I will propose that these problems are solved, or at least alleviated, by the dissociation into space-states, which provides an absolute form of decoherence.

\section{The many-spacetimes interpretation}
\label{s:mwi}

We think that we are forced to suspend the validity of the {\schrod} equation during measurements, because we observe only one of the stories that the {\schrod} equation describes as taking place in parallel.
But could we observe more than one of these stories at once? The {\schrod} equation predicts that even the observers would be ``multiplied'', each of its instances participates in one of the stories and not in the others, of which they are oblivious.

Everett noticed the perfect symmetry of the situation, and saw no reason to favor the story in which one gets an outcome against the competing stories.
He proposed to trust the {\schrod} equation and accept that all stories continue to happen independently, once they are separated \cite{Everett1957RelativeStateFormulationOfQuantumMechanics,Everett1973TheTheoryOfTheUniversalWaveFunction}.
{\schrod} himself proposed earlier, along the same lines, something that he worried may ``seem lunatic'' \cite{SchrodingerMWI1995TheInterpretationOfQuantumMechanics,Barrett1999TheQuantumMechanicsOfMindsAndWorlds,Deutsch2010ApartFromUniverses}.

The result of Everett's realization is the \emph{many-worlds interpretation} (MWI) of quantum mechanics.
But there are still open questions in the MWI.
Various proposals were made to solve them, and some researchers think they are solved. Others think that they cannot be solved and even that the MWI does not deserve to be taken seriously.

In this Section I argue that the space-states approach solves some of these problems, or provides a more natural way to solve them.
This leads to a variant of the many-worlds interpretation, which 
may be called ``the many-space-states interpretation'', but I will call it \emph{the many-spacetimes interpretation} (MSTI).

\subsection{Preferred basis: space-states}
\label{s:preferred-basis}

Let us start with a problem whose solution is the key to solving other problems.

\begin{mwiProblem}[Preferred basis]
\label{mwiProblem:preferred-basis}
In what basis does the branching take place, so that the worlds appear classical at the macro level?
\end{mwiProblem}

Presumably, decoherence makes a preferred basis emerge \cite{Schlosshauer2007DecoherenceAndTheQuantumToClassicalTransition}. However, there has to be more to the preferred basis than that it simply ``emerges''. It was proved that if a preferred basis emerges, either for the entire universe, or for a subsystem, infinitely many others emerge, and they are physically distinct, leading to physically different worlds \cite{Stoica2021SpaceThePreferredBasisCannotUniquelyEmergeFromTheQuantumStructure}.

In the nonrelativistic MWI, it is expected that the preferred basis is related to the positions in the configuration space, or the positions and momenta in the phase space. Accordingly, the branches no longer interfere because they no longer overlap in the configuration space.

The MSTI answer is based on the form of ``absolute decoherence'' provided by the dissociation into space-states enforced by background freedom:
\begin{mstiAnswer}[Preferred basis]
\label{mstiAnswer:preferred-basis}
The dissociation of the state vector automatically selects as the preferred basis the ontic space-states basis.
\end{mstiAnswer}

\subsection{Macro world. Quasi-classicality as classicality}
\label{s:macro}

A related problem is the following
\begin{mwiProblem}[Macro world]
\label{mwiProblem:macro-world}
How does the classical-looking macroscopic world emerge from the wavefunction?
\end{mwiProblem}

Often, Problem \ref{mwiProblem:macro-world} is considered solved by decoherence \cite{Schlosshauer2007DecoherenceAndTheQuantumToClassicalTransition,Zurek2022EmergenceOfTheClassicalWorldFromWithinOurQuantumUniverse,Kiefer2022FromQuantumToClassicalEssaysInHonourOfHDieterZeh}, which appeared in the first place to solve it.

Without denying the importance of decoherence, the process of dissociation makes it stronger, by introducing a notion of ``absolute decoherence'', solving the problem.

\begin{mstiAnswer}[Macro world]
\label{mstiAnswer:macro-world}
Each macro world corresponds to a macrostate, therefore to multiple ontic space-states that belong to the same macrostate, because they are not distinguishable at the macro level.
\end{mstiAnswer}

Multiple space-states can form macrostates (Assumption \ref{as:quasi-classical}). Since each space-state is also quasi-classical, and since they are not distinguished by the macro-projectors, they can account for the macro world.

\subsection{Objects in space}
\label{s:wf-on-3dspace}

Another problem is that the wavefunction is not defined on space, but on the much larger configuration space.
This disturbed {\schrod} \cite{BacciagaluppiValentini2009SolvayConference}, Lorentz (\cite{Przibram1967LettersWaveMechanics}, p. 44), Einstein \cite{Howard1990EinsteinWorriesQM,FineBrown1988ShakyGameEinsteinRealismQT}, Heisenberg, Bohm \cite{Bohm2004CausalityChanceModernPhysics} \etc.
This is true for the wavefunction of any state vector in the total Hilbert space.

\begin{mwiProblem}[Objects in space]
\label{mwiProblem:wavefunction-on-space}
Given that the wavefunction is defined on the high-dimensional configuration space, how do familiar, classical-looking objects localized in space emerge from the wavefunction?
\end{mwiProblem}

The wavefunction, being an element of a representation of the Galilei or the Poincar\'e group \cite{Wigner1959GroupTheoryAndItsApplicationToTheQuantumMechanicsOfAtomicSpectra}, is automatically associated to space or spacetime. Therefore, properly analyzed, it satisfies all expectations of standard geometric objects in space or spacetime \cite{Stoica2021WhyTheWavefunctionAlreadyIsAnObjectOnSpace}.
Moreover, if one thinks that this is not sufficient, and the wavefunction should be expressed as classical-like fields in space or spacetime, this is also possible, albeit in an inaesthetic way that at least serves as a proof of concept \cite{Stoica2019RepresentationOfWavefunctionOn3D}. 

But even if the wavefunction is, in the sense of group theory or as fields, an object in space, in general it does not look like the familiar, classical-looking objects we see. 

Decoherence might lead to branches that look like familiar, classical-looking objects localized in space. Wallace \cite{Wallace2012TheEmergentMultiverseQuantumTheoryEverettInterpretation} thinks that the branches form patterns in the sense of Dennett \cite{Dennett1991RealPatterns}, but are these patterns classical-looking enough to correspond to the observations?

\begin{observation}[{\schrod} atoms in space]
\label{obs:schrod_atom}
Let us start with a simple atom, consisting of $n$ electrons, $n$ protons, and $n'\approx n$ neutrons. Each of the nucleons consist of three quarks, but the things are more complicated, since there is a practically infinite number of virtual quarks and gluons involved, and the interactions between nucleons involve exchange of mesons. But, due to the quarks' large masses compared to the electron's mass, and to their confinement to the short range of the strong interaction, the nucleus of the atom can be very well localized in space.
Let us consider an atom in a state that is a solution of the stationary {\schrod} equation.
We decompose the wavefunction as eigenfunctions of the particle number operator with integer eigenvalue $k$.
These components are very well localized around a point $(x,y,z)^k$ in the configuration space of $k$ particles.
The electrons are spread on a wider region than the nucleus' particles, with an amplitude that vanishes very quickly with the distance from $(x,y,z)^k$. But they are still very strongly peaked around $(x,y,z)^k$.
Since the solution is stationary, the atom is very well localized.
Its position $(x,y,z)$ decouples from the other quantum numbers, so it behaves like a one-particle wavefunction with many ``internal'' degrees of freedom defining a ``rigid object'' consisting mainly of the orbitals.

This picture can form the basis for describing molecules and larger objects as localized objects in space.
The wavefunction, even without decoherence, is a superposition of products of such bound states and free particle states.
Decoherence is expected to decompose the wavefunction into branches that look quasi-classically in this way (Problem \ref{mwiProblem:macro-world}), and nothing else is needed.
\qed
\end{observation}

Maudlin \cite{Maudlin2010CanTheWorldBeOnlyWavefunction,Maudlin2014CriticalStudyDavidWallaceTheEmergentMultiverseQuantumTheoryEverettInterpretation,Maudlin2019PhilosophyofPhysicsQuantumTheory} and Norsen \cite{Norsen2017FoundationsQM} think that Problem \ref{mwiProblem:wavefunction-on-space} is not solved, and that it is hard to solve it even if Problems \ref{mwiProblem:preferred-basis} and \ref{mwiProblem:macro-world} would be.
They contrast this with the pilot-wave theory (PWT) \cite{Bohm1952SuggestedInterpretationOfQuantumMechanicsInTermsOfHiddenVariables}, which includes, along with the wavefunction, point-particles at definite positions in space, and with the Ghirardi-Rimini-Weber (GRW) interpretation \cite{GhirardiRiminiWeber1986GRWInterpretation}, where the wavefunction collapses around definite points of the configuration space, appearing nearly classical.

Their arguments can be seen as relying on the idea that the primitive ontologies of the GRW interpretation -- the \emph{mass density ontology} (GRWm) but especially Bell's \emph{flash ontology} (GRWf) \cite{Bell2004SpeakableFlashOntology} -- and the PWT are very similar to the ontologies of the classical theories. This similarity also seems to help solving the other problems of the PWT and the GRW interpretations.
But for the PWT to work, the wavefunction governing the motion of the particles has to be itself well localized around the points of the configuration space, otherwise the macroscopic objects like chairs and tables, expected to be stable, can explode or transform into completely different configurations.
Also, in the GRW interpretation, to avoid the tails of the wavefunction (which gives the probability density for the spontaneous localization) from allowing the next collapses to lead to abrupt changes in the way the world appears at the macro level, the wavefunction has to decohere as well and its branches have to become well localized in the configuration space.
Therefore, the MWI Problem \ref{mwiProblem:wavefunction-on-space} applies to the PWT and the GRW interpretation as well.
Decoherence and the localized description of the atom from Observation \ref{obs:schrod_atom} are needed by the PWT and the GRW as much as it is needed by the MWI.

But an important lesson that can be learned from their arguments is that classical physics is clearer, and so any interpretation of quantum mechanics that is closer to classical physics has an important advantage.

This suggests the following heuristic rule
\begin{thumbrule}
\label{thumbrule:familiar}
If a solution is considered to work without problems in classical physics, and if it can be applied to an interpretation of quantum mechanics, it should also be considered to work without problems in that interpretation of quantum mechanics.
\end{thumbrule}

We can see that the MSTI Answers \ref{mstiAnswer:preferred-basis}  and \ref{mstiAnswer:macro-world} already align the MWI to this Rule of Thumb, except that the classical objects are fields and not point-particles.
On the other hand, it is hard to see how the beables localized to points in space from the PWT and the GRWf continue to remain as such once we advance these interpretations to quantum fields.

It is therefore desirable to have a solution of Problem \ref{mwiProblem:wavefunction-on-space} along the Rule of Thumb \ref{thumbrule:familiar} as in the PWT and the GRW interpretations, but without adding more structure.
Background freedom automatically makes this possible.

The representation of the wavefunction in terms of fields on space given in \cite{Stoica2019RepresentationOfWavefunctionOn3D} is too dependent on nonrelativistic quantum mechanics, even if it works for the Fock space of quantum field theory too.
In the case of quantum gravity, it works only if the theory is background-dependent.

But the wavefunctional formulation allows for a simpler and more adequate answer:

\begin{observation}
\label{obs:mwiProblem:wavefunction-on-space}
On each space $(\Sigma,\gamma)$, since $\ket{\Psi_{\gamma}}:=\int_{\mc{C}_{M}^{\gamma}}c[\phi_{\gamma}]\ket{\phi_{\gamma}}\De\mu_M^{\gamma}[\phi_{\gamma}]$, it automatically consists of many classical fields $\phi_{\gamma}\in\mc{C}_{M}^{\gamma}$, each of them having attached a complex number $c[\phi_{\gamma}]$.
But since in the polar form $c[\phi_{\gamma}]r[\phi_{\gamma}]e^{i \theta[\phi_{\gamma}]}$ the factor $e^{i \theta[\phi_{\gamma}]}$ can be removed by replacing $\phi_{\gamma}$ with its gauge transformed $e^{i \theta[\phi_{\gamma}]}\phi_{\gamma}$ (\sref{s:real-wavefunctional}), and since $r[\phi_{\gamma}]$ is just a density factor (\sref{s:Born}), what we have is just a collection of classical fields.
States like those containing atoms, molecules, or larger quasi-classical objects are densitized sets of such classical fields, because the Fock space basis can be obtained from the wavefunctional classical basis \cite{Hatfield2018QuantumFieldTheoryOfPointParticlesAndStrings}.
\end{observation}

In addition,
\begin{observation}
\label{obs:mwiProblem:Hamiltonian-on-space}
In the wavefunctional formulation, the Hamiltonian is explicitly local on $\Sigma$ \cite{Hatfield2018QuantumFieldTheoryOfPointParticlesAndStrings}.
\end{observation}

Observations \ref{obs:mwiProblem:wavefunction-on-space} and \ref{obs:mwiProblem:Hamiltonian-on-space} lead to
\begin{mstiAnswer}[Objects in space]
\label{mstiAnswer:wavefunction-on-space}
The space-states consist of classical fields on space.
The dynamics is local.
Densitized sets of such states form worlds as in the MWI only if their dissociation is not manifest at the macro level.
\end{mstiAnswer}

\subsection{Branching asymmetry from Big-Bang symmetry}
\label{s:branching-asymmetry}

Another problem is the following
\begin{mwiProblem}[Branching asymmetry]
\label{mwiProblem:branching-asymmetry}
Why is the branching happening only towards the future, and why do the branches remain separated?
\end{mwiProblem}

This is also often claimed to be solved by decoherence, but since the {\schrod} equation is time-symmetric, without very fine-tuned initial conditions of the universe, decoherence would equally predict branching towards the past. Interference with branches previously separated would affect the probabilities and violate the Born rule.

In the standard framework of the many-worlds interpretation, Wallace acknowledged the problem of branching asymmetry, analyzed it, and concluded that it correlates with the \emph{thermodynamic arrow of time} \cite{Wallace2012TheEmergentMultiverseQuantumTheoryEverettInterpretation}.
For the importance of the thermodynamic arrow of time in relation to branching asymmetry, but also for unforeseen complications, see \cite{Stoica2022DoesQuantumMechanicsRequireConspiracy}.
But we do not have an explanation for the thermodynamic arrow of time either, although the second law of thermodynamics is a well-established empirical fact.

The dissociation into space-states allows us to make progress, by strongly connecting the branching asymmetry with the \emph{cosmological arrow of time}.
This arrow of time points in the time direction in which the universe expands.
The closer the state of the universe is to the Big-Bang, the more homogeneous and isotropic the universe is. Moreover, as the singularity is approached, space contracts.

A possible assumption is that the Big-Bang is a point, the singularity. This would be problematic, since if $\Sigma$ is a lower-dimensional set at $t=0$, we will need to explain how it evolves into a $3$d manifold. To avoid this, one usually assume that the singularity is removed from spacetime, or that quantum gravity eliminates it, so the history starts at $t=\epsilon\searrow 0$, and not at $t=0$.

An alternative option is to embrace the singularity, because the space components of the metric tensor tend to $0$ as $t\searrow 0$, but the topology of space does not contract to a point, it is still the $3$d manifold $(\Sigma,\gamma_{ab}(\x) \equiv 0)$. By avoiding the assumption that the topology derives from distance, we can obtain equations for general relativity that continue to be valid under more general conditions, including at a large class of singularities. For this we need an alternative formulation of semi-Riemannian geometry and Einstein's general relativity, which is equivalent to these ones outside the singularity, but well-defined and free of infinities at the singularity. This was achieved and shown to work in many situations in which non-singular semi-Riemannian geometry is not defined \cite{Stoica2011FLRWBigBangSingularitiesAreWellBehaved,Stoica2012BeyondTheFLRWBigBangSingularity,Stoica2013SingularGeneralRelativityPhDThesis}.
Moreover, when it is valid at the Big-Bang, it automatically satisfies Penrose's \emph{Weyl curvature hypothesis}, whose motivation was in the first place to connect the cosmological and the thermodynamic arrows of time \cite{Penrose1979SingularitiesAndTimeAsymmetry,Stoica2012OnTheWeylCurvatureHypothesis}.

Then, there is only one possible space-state at the Big-Bang, $(\Sigma,\gamma_{ab}(\x) \equiv 0)$.
The condition of validity of \emph{singular general relativity} from \cite{Stoica2013SingularGeneralRelativityPhDThesis} also requires that the matter fields are homogeneous at the Big-Bang \cite{Stoica2014GaugeTheoryAtSingularities}.

As $t\searrow 0$ the system may be chaotic, as in the Mixmaster model \cite{Misner1969MixmasterUniverse} or the Belinski–Khalatnikov–Lifshitz model \cite{BKL1970OscillatoryApproachToASingularPointInTheRelativisticCosmology}.
Then, while at the singularity there is still only one possible space-state, it can be approached in different ways as $t\searrow 0$.
However, the limit $\gamma\to 0$ forces the solutions to depend on a small number of parameters as they converge to the unique space $(\Sigma,\gamma_{ab}(\x) \equiv 0)$.

This severe constraint of the initial conditions for $(\Sigma,\gamma_{ab})$ implies that the branching structure of the wavefunctional is extremely asymmetric in time.
This suggests a possible reason why, at macro scales, branching happens only towards the future.

\begin{mstiAnswer}[Branching asymmetry]
\label{mstiAnswer:branching-asymmetry}
Branching happens only towards the future because at the Big-Bang the space-states have a very small number of degrees of freedom and all converge to the same initial space-state $(\Sigma,\gamma_{ab}(\x) \equiv 0)$.
\end{mstiAnswer}

This answer is of course incomplete. We do not know why the initial state had to be the Big-Bang, and it is not even sure that there was a singularity, many researchers think that quantum gravity will be able to remove it. But singular general relativity suggests that singularities are not necessarily ill-behaved, they can even solve notorious problems in quantum gravity \cite{Stoica2012MetricDimensionalReductionAtSingularitiesWithImplicationsToQuantumGravity,Stoica2018RevisitingBlackHoleEntropyAndTheInformationParadox}.

\subsection{Probabilities from continuity}
\label{s:probabilities-continuity}

When a quantum measurement is made, the probability to obtain a certain outcome is given by the Born rule to be the square of the projection of the state vector. Different outcomes may therefore have different probabilities.
However, in the MWI, there is only one branch for each of these outcomes.
A direct counting argument implies that all outcomes should be obtained with the same probability, contrary to the Born rule.

\begin{mwiProblem}[Probabilities]
\label{mwiProblem:probabilities}
Why are the probabilities proportional to the squared amplitudes of the branches?
\end{mwiProblem}

There are various proposed solutions, based on many-minds
\cite{AlbertLower1988InterpretingMWI_ManyMinds}, decision theory \cite{Deutsch1999QuantumTheoryOfProbabilityAndDecision,Wallace2002QuantumProbabilitiesAndDecisionRevisited}, measure of existence \cite{Vaidman2012ProbabilityInMWI} \etc. For a review see \cite{Vaidman2020DerivationsOfTheBornRule}. 
Proposals that somehow the amplitude of a branch yields probability have merits and they lead to interesting insights into the nature of probability \cite{Wallace2012TheEmergentMultiverseQuantumTheoryEverettInterpretation}.
But if probabilities could be obtained in the old-fashioned way, for example by branch counting of a more refined branching structure (Saunders advocates this \cite{Saunders2021BranchCountingInTheEverettInterpretationOfQuantumMechanics})
or ideally as \emph{the ratio of the number of favorable outcomes to the total number of possible outcomes}, the result would be more palatable, without necessarily contradicting other proposals.

These proposals are well justified, but it may help to have a solution based on microstate or micro-branch counting, according to the Rule of Thumb \ref{thumbrule:familiar}.
A micro-branch counting method would, preferably, rely on an absolute notion of micro-branches.
The MSTI does just this:
\begin{mstiAnswer}[Probabilities]
\label{mstiAnswer:probabilities}
``Counting'' space-states allowed to have an inhomogeneous density gives probabilities proportional to the squared amplitudes.
Taking each space-state into account is justified by the fact that only those states have local beables, see Principle \ref{pp::worlds} and \sref{s:ontology}.
\end{mstiAnswer}

This may provide a concrete realization of other proposals, without necessarily rejecting them.

\subsection{Real wavefunction}
\label{s:wavefunction-gauge}

The Rule of Thumb \ref{thumbrule:familiar} also suggests the following
\begin{mwiProblem}[Real-number-based probabilities]
\label{mwiProblem:real-numbers}
It is true that the norm of the (complex) wavefunction is real. But is there a deeper reason why we get real probabilities?
\end{mwiProblem}

The MSTI suggests a solution according to the Rule of Thumb \ref{thumbrule:familiar} for this too:
\begin{mstiAnswer}[Real-number-based probabilities]
\label{mstiAnswer:real-numbers}
The wavefunctional is real, and the phases only represent a global $\U(1)$ gauge choice for the classical fields in the space-states.
The real coefficients are the densities of the space-states composing the wavefunctional.
\end{mstiAnswer}

\subsection{Ontology}
\label{s:ontology}

Sometimes it is said that the wavefunction lacks local beables (in space). 

\begin{mwiProblem}[Ontology]
\label{mwiProblem:ontology}
What is the ontology of the MWI?
What are the local beables?
\end{mwiProblem}

A primitive ontology similar to that of the GRWm (see \sref{s:wf-on-3dspace}) can work for the MWI \cite{AGTZ2010ManyWorldsScrodinger}, by using as local beables the mass or charge density, solution originally proposed by {\schrod} to interpret the wavefunction. But it can be argued that this adds new ontology besides the wavefunction, and that the same can be said if we take the center of mass of the atoms as the local beables in the MWI.

Some researchers consider that the abstract state vector and the Hamiltonian are sufficient to specify the ontology of the MWI, and from it one can derive an essentially unique space, the tensor product structure, the preferred basis, and all there is to be known about the universe \cite{CarrollSingh2019MadDogEverettianism}.
But if any of these structures can be derived from the state vector and the Hamiltonian, infinitely many other physically distinct solutions exist \cite{Stoica2021SpaceThePreferredBasisCannotUniquelyEmergeFromTheQuantumStructure}.
So a preferred basis or another structure should exist.

And indeed, other researchers consider that not merely the abstract state vector is needed, but the wavefunction, \ie the state vector expressed in space, and this is sufficient to specify the complete ontology \cite{Vaidman2016AllIsPsi,SEP-Vaidman2018MWI}.

Despite this, and despite the arguments from \cite{Stoica2021WhyTheWavefunctionAlreadyIsAnObjectOnSpace} and the local representation of the wavefunction from \cite{Stoica2019RepresentationOfWavefunctionOn3D}, researchers like Maudlin \cite{Maudlin2010CanTheWorldBeOnlyWavefunction,Maudlin2014CriticalStudyDavidWallaceTheEmergentMultiverseQuantumTheoryEverettInterpretation,Maudlin2019PhilosophyofPhysicsQuantumTheory} and Norsen \cite{Norsen2017FoundationsQM} consider that the MWI does not have a primitive ontology (in terms of local beables).

But in every micro-world in the MSTI there are local beables, just like in classical physics.
\begin{mstiAnswer}[Ontology]
\label{mstiAnswer:ontology}
Although the wavefunctional is all there is \cite{Vaidman2016AllIsPsi}, the space-states have a privileged ontic status, so we will call them ontic states. This is necessary, because the ontology cannot be the abstract state vector \cite{Stoica2021SpaceThePreferredBasisCannotUniquelyEmergeFromTheQuantumStructure}, but the wavefunction, which presuposes a preferred basis.
Each of the ontic states consists of a space $(\Sigma,\gamma)$, on which classical fields $\phi_{\gamma}$ are defined. Each ontic state appears at most once in the composition of the wavefunctional, with a nonuniform density.
The distribution gives the real wavefunctional, and a global gauge gives the phase of each term in the wavefunctional. The local beables are the classical fields $\phi_{\gamma}$ and $\gamma$ defined on the $3$d manifold $\Sigma$, so they are defined only for space-states.
The space-states, having definite local beables, correspond to (micro-)worlds. This justifies counting them to obtain the probabilities.
\end{mstiAnswer}

Since local beables exist as classical fields (including the metric $\gamma$), the Rule of Thumb \ref{thumbrule:familiar} was followed.
What can be more classical than the classical itself?

\section{Discussion}
\label{s:discussion}

It is uncommon to use the wavefunctional formulation of quantum field theory in the interpretation of quantum mechanics. It is more common to take nonrelativistic quantum mechanics as a benchmark for these interpretations. But the wavefunctional formulation is natural too, if not even more natural, and it is more realistic.

\begin{observation}
\label{obs:measurement}
When we perform a quantum measurement of a smaller system, we never observe directly its state, only the pointer state of the apparatus, which is macroscopic.
A measuring device is dedicated to a particular location and type of quantum field (or subsystem in general), not to a particular particle (or subsystem).
The result of any measurement translates into a change in the macrostate of the universe.
All these are described adequately by the wavefunctional of the entire universe.
\end{observation}

Wheeler and Everett considered the MWI as \emph{the} interpretation of quantum mechanics that is suitable for quantum gravity \cite{Byrne2010TheManyWorldsOfHughEverettIII,Barrett2012EverettInterpretation}.
According to DeWitt \cite{Dewitt1967QuantumTheoryOfGravityI_TheCanonicalTheory}, p. 1141:
\begin{quote}
Everett's view of the world is a very natural one to adopt in the quantum theory of gravity, where one is accustomed to speak without embarrassment of the `wave function of the universe.' It is possible that Everett's view is not only natural but essential.
\end{quote}

Here, we have seen that background free quantum gravity solves some foundational problems of quantum mechanics, and especially of the MWI. It even suggests a version of the MWI, which is the MSTI, as the more natural interpretation of quantum mechanics. The relation between quantum gravity and the MWI is therefore reciprocal.

Finally, I argued that the MSTI solves some of the main problems of standard quantum mechanics and the MWI. 
The strategy to make this interpretation more palatable was to highlight similarities with classical physics, based on the Rule of Thumb \ref{thumbrule:familiar}.
It turns out that, except for the existence of a multiplicity of worlds, the MSTI is a more classical-like version of the MWI, with respect to the appearance of classicality, the existence of local beables, the probabilities, and even the understanding of the complex numbers inherent to the theory.

\textbf{Acknowledgement} The author thanks Simon Saunders for valuable comments and suggestions offered to a previous version of the manuscript. Nevertheless, the author bears full responsibility for the article.


\end{document}